\def\year{2016}\relax
\documentclass[letterpaper]{article}
\usepackage{aaai16}
\usepackage{times}
\usepackage{helvet}
\usepackage{courier}
\usepackage{graphicx}
\usepackage{amsmath}
\usepackage{amssymb}
\usepackage{algorithm}
\usepackage{multirow} 
\usepackage{algorithmic}
\usepackage{booktabs,caption,fixltx2e}
\usepackage[flushleft]{threeparttable}
\usepackage{subfig}
\usepackage{threeparttable}
\renewcommand{\vec}[1]{\mathbf{#1}}
\newtheorem{theorem}{Theorem}

\DeclareMathOperator*{\argmin}{arg\,min}
\begin{document}
%
\title{Top-N Recommender System via Matrix Completion}
\author{Zhao Kang\quad Chong Peng\quad Qiang Cheng\\
Department of Computer Science, Southern Illinois University, Carbondale, IL 62901, USA\\
\{Zhao.Kang, pchong, qcheng\}@siu.edu\\
}
\maketitle
\begin{abstract}
Top-N recommender systems have been investigated widely both in industry and academia. However, the recommendation quality is far from satisfactory. In this paper, we propose a simple yet promising algorithm. We fill the user-item matrix based on a low-rank assumption and simultaneously keep the original information. To do that, a nonconvex rank relaxation rather than the nuclear norm is adopted to provide a better rank approximation and an efficient optimization strategy is designed. A comprehensive set of experiments on real datasets demonstrates that our method pushes the accuracy of Top-N recommendation to a new level.

\end{abstract}

\section{Introduction}
The growth of online markets has made it increasingly difficult for people to find items which are interesting and useful to them. Top-N recommender systems have been widely adopted by the majority of e-commerce web sites to recommend size-$N$ ranked lists of items that best fit customers' personal tastes and special needs \cite{linden2003amazon}. It works by estimating a consumer's  response for new items, based on historical information, and suggesting to the consumer novel items for which the predicted response is high. In general, historical information can be obtained explicitly, for example, through ratings/reviews, or implicitly, from purchase history or access patterns \cite{desrosiers2011comprehensive}.

Over the past decade, a variety of approaches have been proposed for Top-N recommender systems \cite{ricci2011introduction}. They can be roughly divided into three categories: 
neighborhood-based collaborative filtering, model-based collaborative filtering, and ranking-based methods. The general principle of neighborhood-based methods is to identify the similarities among users/items \cite{deshpande2004item}. For example, item-based k-nearest-neighbor (ItemKNN) collaborative filtering methods first identify a set of similar items for each of the items that the consumer has purchased, and then recommend Top-N items based on those similar items. However, they suffer from low accuracy since they employ few item characteristics.

Model-based methods build a model and then generate recommendations. For instance, the widely studied matrix factorization (MF) methods employ the user-item similarities in their latent space to extract the user-item purchase patterns. Pure singular-value-decomposition-based (PureSVD) matrix factorization method \cite{cremonesi2010performance} characterizes users and items by the most principal singular vectors of the user-item matrix. A weighted regularized matrix factorization (WRMF) \cite{pan2008one,hu2008collaborative} method applies a weighting matrix to differentiate the contributions from observed purchase/rating activities and unobserved ones.

The third category of methods rely on ranking/retrieval criteria. Here, Top-N recommendation is treated as a ranking problem. A Bayesian personalized ranking (BPR) \cite{rendle2009bpr} criterion, which is the maximum posterior estimator from a Bayesian analysis, is used to measure the difference between the rankings of user-purchased items and the rest items. BPR can be combined with ItemKNN (BPRkNN) and MF method (BPRMF). One common drawback of these approaches lies in low recommendation quality.

Recently, a novel Top-N recommendation method SLIM \cite{ning2011slim} has been proposed. From user-item matrix $X$ of size $m\times n$, it first learns a sparse aggregation coefficient matrix $W\in\mathcal{R}_{+}^{n\times n}$ by encoding each item as a linear combination of all other items and solving an $l_1$-norm and $l_2$-norm regularized optimization problem. Each entry $w_{ij}$ describes the similarity between item $i$ and $j$. SLIM has obtained better recommendation accuracy than the other state-of-the-art methods. However, SLIM can only capture relations between items that are co-purchased/co-rated by at least one user, while an intrinsic characteristic of recommender systems is sparsity due to the fact that users typically rate only a small portion of the available items. 

To overcome the above limitation, LorSLIM \cite{cheng2014lorslim} has also imposed a low-rank constraint on $W$. It solves the following problem:
\begin{equation*}
\begin{split}
&\min_W \frac{1}{2}\|X-XW\|_F^2+\frac{\beta}{2}\|W\|_F^2+\lambda \|W\|_1+\gamma\|W\|_*    \\
&s.t.\quad W\ge 0, \quad diag(W)=0,   
\end{split}
\label{lorslim}
\end{equation*}
where $\|W\|_*$ is the nuclear norm of $W$, defined as the sum of its singular values. Low-rank structure is motivated by the fact that a few latent variables from $F$ that explain items' features in factor model $W\approx FF^T$ are of low rank. After obtaining $W$, the recommendation score for user $i$ about an un-purchased/-rated item $j$ is $\hat{x}_{ij}=\vec{x}_i^T\vec{w}_j$, where $x_{ij}=0$, $\vec{x}_i^T$ is the $i$-th row of $X$, and $\vec{w}_j$ is the $j$-th column of $W$. Thus $\hat{X}=XW$. 
LorSLIM can model the relations between items even on sparse datasets and thus improves the performance. 

To further boost the accuracy of Top-N recommender systems, we first fill the missing ratings by solving a nonconvex optimization problem, based on the assumption that the user' ratings are affected by only a few factors and the resulting rating matrix should be of low rank \cite{lee2014local}, and then make the Top-N recommendation. This is different from previous approaches: Middle values of the rating ranges, or the average user or item ratings are commonly utilized to fill out the missing ratings \cite{breese1998empirical,deshpande2004item}; a more reliable approach utilizes content information \cite{melville2002content,li2004combining,degemmis2007content}, for example, the missing ratings are provided by autonomous agents called filterbots \cite{good1999combining}, which rate items based on some specific characteristics of their content; a low rank matrix factorization approach seeks to approximate $X$ by a multiplication of low rank factors \cite{yu2009fast}. Experimental results demonstrate the superior recommendation quality of our approach.

Due to the inherent computational complexity of rank problems, the non-convex rank function is often relaxed to its convex surrogate, i.e. the nuclear norm \cite{candes2009exact,recht2008necessary}. However, this substitution is not always valid and can lead to a biased solution \cite{shi2011limitations,kangcikm2015robust}. Matrix completion with nuclear norm regularization can be significantly hurt when entries of the matrix are sampled non-uniformly \cite{srebro2010collaborative}. Nonconvex rank approximation has received significant attention \cite{zhong2015nonconvex,kangICDM}. Thus we use log-determinant ($logdet$) function to approximate the rank function and design an effective optimization algorithm. 

\section{Problem Formulation}
The incomplete user-item purchases/ratings matrix is denoted as $M$ of size $m \times n$. $M_{ij}$ is 1 or a positive value if user $i$ has ever purchased/rated item $j$; otherwise it is 0. Our goal is to reconstruct a full matrix $X$, which is supposed to be low-rank.  
Consider the following matrix completion problem:
\begin{equation}
\begin{split}
\min_X logdet ((X^TX)^{1/2}+I)   \\
s.t.\quad X_{ij}=M_{ij},\quad (i,j)\in \Omega,  
\end{split}
\label{originalprob}
\end{equation}
where $\Omega$ is the set of locations corresponding to the observed entries and $I\in \mathcal{R}^{n\times n}$ is an identity matrix. It is easy to show that $logdet((X^TX)^{1/2}+I)\leq\|X\|_*$, i.e., $logdet$ is a tighter rank approximation function than the nuclear norm. $logdet$ also helps mitigate another inherent disadvantage of the nuclear norm, i.e., the imbalanced penalization of different singular values \cite{kang2015robust}. Previously $logdet(X+\delta I)$ was suggested to restrict the rank of positive semidefinite matrix $X$ \cite{fazel2003log}, which is not guaranteed for more general $X$, and also $\delta$ is required to be small, which leads to significantly biased approximation for small singular values. Compared to some other nonconvex relaxations in the literature \cite{lu2014generalized}, our formulation enjoys the simplicity and efficacy. 

\section{Methodology}
 Considering that the user-item matrix is often nonnegative, we add nonnegative constraint $X\ge 0$, i.e., element-wise positivity, for easy interpretation of the representation. Let $\mathcal{P}_\Omega$ be the orthogonal projection operator onto the span of matrices vanishing outside of $\Omega$ (i.e., $\Omega^c$) so that 
\begin{eqnarray*}
(\mathcal{P}_\Omega(X))_{ij}=\left\{
\begin{array}{ll} X_{ij}, & \mbox{if $X_{ij}\in\Omega$};\\
0, & \mbox{if $X_{ij}\in\Omega^c$.}
\end{array}\right.
\end{eqnarray*} 
Problem (\ref{originalprob}) can be reformulated as 
\begin{equation}
\begin{split}
\min_X logdet ((X^TX)^{1/2}+I)+l_{\mathcal{R}_+}(X)   \\
s.t.\quad \mathcal{P}_\Omega(X)=\mathcal{P}_\Omega(M),  
\end{split}
\end{equation}
where $l_{\mathcal{R}_+}$ is the indicator function, defined element-wisely as
\begin{eqnarray*}
\label{error21}
l_{\mathcal{R}_+}(x)=\left\{
\begin{array}{ll} 0, & \mbox{if $x\geq 0$};\\
+\infty, & \mbox{otherwise.}
\end{array}\right.
\end{eqnarray*}
Notice that this is a nonconvex optimization problem, which is not easy to solve in general. Here we develop an effective optimization strategy based on augmented Lagrangian
multiplier (ALM) method. By introducing an auxiliary variable $Y$, it has the following equivalent form
\begin{equation}
\label{secondprob}
\begin{split}
&\min_{X, Y} logdet ((X^TX)^{1/2}+I)+l_{\mathcal{R}_+}(Y)   \\
&s.t.\quad \mathcal{P}_\Omega(X)=\mathcal{P}_\Omega(M),\quad X=Y,  
\end{split}
\end{equation}
which has an augmented Lagrangian function of the form
\begin{equation}
\begin{split}
\mathcal{L}(X,Y,Z)= logdet ((X^TX)^{1/2}+I)+l_{\mathcal{R}_+}(Y)+\\
\frac{\mu}{2}\|X-Y+\frac{Z}{\mu}\|_F^2
\quad s.t.\quad \mathcal{P}_\Omega(X)=\mathcal{P}_\Omega(M),  
\end{split}
\end{equation}
where $Z$ is a Lagrange multiplier and $\mu>0$ is a penalty parameter.



Then, we can apply the alternating minimization idea to update $X$, $Y$, i.e., updating one of the two variables with the other fixed.

Given the current point $X^t$, $Y^t$,  $Z^t$, we update $X^{t+1}$ by solving 
\begin{equation}
\begin{split}
X^{t+1}&=\argmin_{X} logdet ((X^TX)^{1/2}+I)+\\
&\frac{\mu^t}{2}\|X-Y^t+\frac{Z^t}{\mu^t}\|_F^2
\end{split}
\label{noncvx}
\end{equation}
This can be converted to scalar minimization problems due to the following theorem \cite{kang2015logdet}.  
\begin{theorem}
\label{thm}
If $F(Z)$ is a unitarily invariant function and SVD of $A$ is $A = U \Sigma_A V^T$, then the optimal solution to the following problem 
\begin{equation}
\min_{Z}F(Z)+\frac{\beta}{2}\|Z-A\|_F^{2}
\label{eq:Zobj}
\end{equation}
 is $Z^*$ with SVD $U\Sigma_Z^* V^T$, where $\Sigma_Z^* = diag\left(\sigma^*\right)$; moreover, $F(Z) = f \circ \sigma(Z)$, where $\sigma(Z)$ is the vector of nonincreasing  singular values of $Z$, 
 then $\sigma^*$ is obtained by using the Moreau-Yosida proximity operator
$\sigma^* = prox_{f, \beta} (\sigma_{A})$, where $\sigma_A := diag(\Sigma_A)$, and 
\begin{equation}
\label{scalar}
prox_{f, \beta} (\sigma_A) := \argmin_{\sigma\ge 0} f(\sigma) + \frac{\beta}{2}\|\sigma - \sigma_A\|_2^2.
\end{equation}
\end{theorem}
According to the first-order optimality condition, the gradient of the objective function of  (\ref{scalar}) with respect to each singular value should vanish.  
For $logdet$ function, we have 
\begin{equation}
\label{svdf}
\frac{1}{1+\sigma_i}+\beta (\sigma_{i}-\sigma_{i,A})=0\hspace{0.1cm} s.t. \hspace{0.1cm}\sigma_i \ge  0.
\end{equation} 
The above equation is quadratic and gives two roots. If $\sigma_{i,A}=0$, the minimizer $\sigma_i^*$ will be 0; otherwise, there exists a unique minimizer. Finally, we obtain the update of $X$ variable with
\begin{equation}
X^{t+1}=U diag(\sigma^{*}) V^T.
\end{equation}
Then we fix the values at the observed entries and obtain
\begin{equation}
\label{solveJ}
X^{t+1}=\mathcal{P}_{\Omega^c}(X^{t+1})+\mathcal{P}_\Omega(M).
\end{equation}

To update $Y$, we need to solve
\begin{equation}
\min_Y \quad l_{\mathcal{R}_+}(Y)+\frac{\mu^t}{2}\|X^{t+1}-Y+\frac{Z^t}{\mu^t}\|_F^2,
\end{equation}
which yields the updating rule
\begin{equation}
\label{solveY}
Y^{t+1}= \textrm{max}(X^{t+1}+Z^t/\mu^t,0).
\end{equation}
Here $\max(\cdot)$ is an element-wise operator. The complete procedure is outlined in Algorithm 1.
\begin{algorithm}[tb]
\small
   \caption{Solve (\ref{secondprob})}
   \label{alg:rankminimization}
  {\bfseries Input:} Original imcomplete data matrix $M_\Omega\in \mathbf{\mathcal{R}}^{m\times n}$, parameters  $\mu^0>0$, $\gamma>1$.\\
{\bfseries Initialize:} $Y=\mathcal{P}_\Omega(M)$, $Z=0$.\\
  {\bfseries REPEAT}
\begin{algorithmic}[1]
   \STATE Obtain $X$ through (\ref{solveJ}).
   \STATE Update $Y$ as (\ref{solveY}).
\STATE Update the Lagrangian multipliers $Z$ by
\begin{align*}
Z^{t+1}&=Z^{t}+\mu^t(X^{t+1}-Y^{t+1}).
\end{align*}
\STATE Update the parameter $\mu^t$ by $\mu^{t+1}=\gamma\mu^t$.
\end{algorithmic}
\textbf{ UNTIL} {stopping criterion is met.}
\end{algorithm}

To use the estimated matrix $\hat{X}$ to make recommendation for user $i$, we just sort $i$'s non-purchased/-rated items based on their scores in decreasing order and recommend the Top-N items. 

\section{Experimental Evaluation}
\subsection{Datasets}
\begin{table}[ht]
\caption{The datasets used in evaluation}
\label{tab:data}
\begin{center}
\tiny
\resizebox{.45\textwidth}{!}{
\begin{tabular}{llllllll}
\multicolumn{1}{c}{dataset}  &\multicolumn{1}{c}{\#users} &\multicolumn{1}{c}{ \#items}  &\multicolumn{1}{c}{\#trns} &\multicolumn{1}{c}{rsize}  &\multicolumn{1}{c}{csize}&\multicolumn{1}{c}{ density}&\multicolumn{1}{c}{ratings}\\
\hline\hline \\
Delicious&1300&4516&17550&13.50&3.89&0.29\%&-\\
 lastfm&8813&6038&332486&37.7&55.07&0.62\%&- \\
BX&4186&7733&182057&43.49&23.54&0.56\%&- \\
\hline \\
ML100K &943&1682&100000&106.04&59.45&6.30\%&1-10\\
Netflix&6769&7026&116537&17.21&16.59&0.24\%&1-5 \\
Yahoo&7635&5252&212772&27.87&40.51&0.53\% &1-5 \\
\hline
\end{tabular}}
   \begin{tablenotes}
      \tiny
      \item The ``\#users", ``\#items", ``\#trns" columns show the number of users, number of items and number of transactions, respectively, in each dataset. The ``rsize" and ``csize" columns are the average number of ratings for each user and on each item (i.e., row density and column density of the user-item matrix), respectively, in each dataset. Column corresponding to ``density" shows the density of each dataset (i.e., density=\#trns/(\#users$\times$\#items)). The ``ratings" column is the rating range of each dataset with granularity 1. 
    \end{tablenotes}
\end{center}
\end{table}
We evaluate the performance of our method on six different real datasets whose characteristics are summarized in Table \ref{tab:data}. These datasets are from different sources and at different sparsity levels. They can be broadly categorized into two classes.

The first class includes Delicious, lastfm and BX. These three datasets have only implicit feedback (e.g., listening history), i.e., they are represented by binary matrices. In particular, Delicious was derived from the bookmarking and tagging information from a set of 2$K$ users from Delicious social bookmarking system\footnote{http://www.delicious.com} such that each URL was bookmarked by at least 3 users. Lastfm corresponds to music artist listening information which was obtained from the last.fm online music system\footnote{ http://www.last.fm }, in which each music artist was listened to by at least 10 users and each user listened to at least 5 artists. BX is a subset from the Book-Crossing dataset\footnote{http://www.informatik.uni-freiburg.de/~cziegler/BX/} such that only implicit interactions were contained and each book was read by at least 10 users. 

The second class contains ML100K, Netflix and Yahoo. All these datasets contain multi-value ratings. Specifically, the ML100K dataset corresponds to movie ratings and is a subset of the MovieLens research project\footnote{http://grouplens.org/datasets/movielens/}. The Netflix is a subset of data extracted from Netflix Prize dataset\footnote{http://www.netflixprize.com/} and each user rated at least 10 movies. The Yahoo dataset is a subset obtained from Yahoo!Movies user ratings\footnote{http://webscope.sandbox.yahoo.com/catalog.php?datatype=r}. In this dataset, each user rated at least 5 movies and each movie was rated by at least 3 users.

\begin{table*}[ht]
\begin{center}
\begin{threeparttable}
\caption{Comparison of Top-N recommendation algorithms}
\label{tab:comp}
\tiny
\begin{tabular}{llllllllllllll}
\hline
\multirow{2}{*}{method} &
\multicolumn{6}{c}{Delicious} &
\multicolumn{1}{c}{}&
\multicolumn{6}{c}{lastfm} \\
\cline{2-7} \cline{9-14} 
  & \multicolumn{4}{c}{params} &\multicolumn{1}{c}{HR}  & \multicolumn{1}{c}{ARHR} &\multicolumn{1}{c}{}&\multicolumn{4}{c}{params} &\multicolumn{1}{c}{HR}  & \multicolumn{1}{c}{ARHR} \\ 
\hline
ItemKNN&300&-&-&-&0.300&0.179&&100&-&-&-&0.125&0.075\\
PureSVD&1000&10&-&-&0.285&0.172&&200&10&-&-&0.134&0.078\\
WRMF&250&5&-&-&0.330&0.198&&100&3&-&-&0.138&0.078\\
BPRKNN&1e-4&0.01&-&-&0.326&0.187&&1e-4&0.01&-&-&0.145&0.083\\
BPRMF&300&0.1&-&-&0.335&0.183&&100&0.1&-&-&0.129&0.073\\
SLIM&10&1&-&-&0.343&0.213&&5&0.5&-&-&0.141&0.082\\
LorSLIM&10&1&3&3&0.360&0.227&&5&1&3&3&0.187&0.105\\
Our&250&4&-&-&\bf{0.382}&\bf{0.241}&&0.03&1.5&-&-&\bf{0.206}&\bf{0.113}\\
\hline\hline
\multirow{2}{*}{method} &
\multicolumn{6}{c}{BX} &
\multicolumn{1}{c}{}&
\multicolumn{6}{c}{ML100K} \\
\cline{2-7} \cline{9-14} 
  & \multicolumn{4}{c}{params} &\multicolumn{1}{c}{HR}  & \multicolumn{1}{c}{ARHR} &\multicolumn{1}{c}{}&\multicolumn{4}{c}{params} &\multicolumn{1}{c}{HR}  & \multicolumn{1}{c}{ARHR} \\ 
\hline
ItemKNN&400&-&-&-&0.045&0.026&&10&-&-&-&0.287&0.124\\
PureSVD&3000&10&-&-&0.043&0.023&&100&10&-&-&0.324&0.132\\
WRMF&400&5&-&-&0.047&0.027&&50&1&-&-&0.327&0.133\\
BPRKNN&1e-3&0.01&-&-&0.047&0.028&&2e-4&1e-4&-&-&0.359&0.150\\
BPRMF&400&0.1&-&-&0.048&0.027&&200&0.1&-&-&0.330&0.135\\
SLIM&20&0.5&-&-&0.050&0.029&&2&2&-&-&0.343&0.147\\
LorSLIM&50&0.5&2&3&0.052&0.031&&10&8&5&3&0.397&0.207\\
Our&1.2e-3&1.3&-&-&\bf{0.065}&\bf{0.043}&&6e-3&2.5&-&-&\bf{0.428}&\bf{0.215}\\
\hline\hline
\multirow{2}{*}{method} &
\multicolumn{6}{c}{Netflix} &
\multicolumn{1}{c}{}&
\multicolumn{6}{c}{Yahoo} \\
\cline{2-7} \cline{9-14} 
  & \multicolumn{4}{c}{params} &\multicolumn{1}{c}{HR}  & \multicolumn{1}{c}{ARHR} &\multicolumn{1}{c}{}&\multicolumn{4}{c}{params} &\multicolumn{1}{c}{HR}  & \multicolumn{1}{c}{ARHR} \\ 
\hline
ItemKNN&200&-&-&-&0.156&0.085&&300&-&-&-&0.318&0.185\\
PureSVD&500&10&-&-&0.158&0.089&&2000&10&-&-&0.210&0.118\\
WRMF&300&5&-&-&0.172&0.095&&100&4&-&-&0.250&0.128\\
BPRKNN&2e-3&0.01&-&-&0.165&0.090&&0.02&1e-3&-&-&0.310&0.182\\
BPRMF&300&0.1&-&-&0.140&0.072&&300&0.1&-&-&0.308&0.180\\
SLIM&5&1.0&-&-&0.173&0.098&&10&1&-&-&0.320&0.187\\
LorSLIM&10&3&5&3&0.196&0.111&&10&1&2&3&0.334&0.191\\
Our&0.015&1.2&-&-&\bf{0.226}&\bf{0.127}&&5e-3&1.1&-&&\bf{0.367}&\bf{0.218}\\
\hline
\end{tabular}

   \begin{tablenotes}

      \item The parameters for each method are as follows: ItemKNN: the number of neighbors $k$; PureSVD: the number of singular values and the number of iterations during SVD; WRMF: the dimension of the latent space and the weight on purchases; BPRKNN: the learning rate and regularization parameter $\lambda$; BPRMF: the dimension of the latent space and learning rate; SLIM: the $l_2$-norm regularization parameter $\beta$ and the $l_1$-norm regularization parameter
$\lambda$; LorSLIM: the $l_2$-norm regularization parameter $\beta$, the $l_1$-norm regularization parameter $\lambda$, the nuclear norm regularization parameter $z$ and the auxiliary parameter $\rho$. Our: auxiliary parameters $\mu^0$ and $\gamma$. N in this table is 10. Bold numbers are the best performance in terms of HR and ARHR for each dataset. 

 \end{tablenotes}
\end{threeparttable}
\end{center}
\end{table*}
\subsection{Evaluation Methodology}
We employ 5-fold Cross-Validation to demonstrate the efficacy of our proposed approach. For each run, each of the datasets is split into training and test sets by randomly selecting one of the non-zero entries for each user to be part of the test set\footnote{We use the same data as in \cite{cheng2014lorslim}, with partitioned datasets kindly provided by Yao Cheng.}. The training set is used to train a model, then a size-N ranked list of recommended items for each user is generated. The evaluation of the model is conducted by comparing the recommendation list of each user and the item of that user in the test set. For the following results reported in this paper, $N$ is equal to 10.  

Top-N recommendation is more like a ranking problem rather than a prediction task. The recommendation quality is measured by the hit-rate (HR) and the average reciprocal hit-rank (ARHR) \cite{deshpande2004item}. They directly measure the performance of the model on the ground truth data, i.e., what users have already provided feedback for. As pointed out in \cite{ning2011slim}, they are the most direct and meaningful measures in Top-N recommendation scenarios. HR is defined as
\begin{equation}
HR=\frac{\#\textrm{hits}}{\#\textrm{users}},
\end{equation}
where \#hits is the number of users whose item in the test set is recommended (i.e., hit) in the size-N recommendation list, and \#users is the total number of users. An HR value of 1.0 indicates that the algorithm is able to always recommend the hidden item, whereas an HR value of 0.0 denotes that the algorithm is not able to recommend any of the hidden items. 

A drawback of HR is that it treats all hits equally regardless of where they appear in the Top-N list. ARHR addresses it by rewarding each hit based on where it occurs in the Top-N list, which is defined as follows:
\begin{equation}
ARHR=\frac{1}{\#\textrm{users}}\sum_{i=1}^{\#\textrm{hits}}\frac{1}{p_i},
\end{equation}
where $p_i$ is the position of the test item in the ranked Top-N list for the $i$-th hit. That is, hits that occur earlier in the ranked list are weighted higher than those occur later, and thus ARHR measures how strongly an item is recommended. The highest value of ARHR is equal to the hit-rate and occurs when all the hits occur in the first position, whereas the lowest value is equal to HR/N when all the hits occur in the last position of the list.

\subsection{Comparison Algorithms}
We compare the performance of the proposed method\footnote{The implementation of our method is available at: https://github.com/sckangz/recom\_mc.} with seven other state-of-the-art Top-N recommendation algorithms, including the item neighborhood-based collaborative filtering method ItemKNN \cite{deshpande2004item}, two MF-based methods PureSVD \cite{cremonesi2010performance} and WRMF \cite{hu2008collaborative}, two ranking/retrieval criteria based methods BPRMF and BPRKNN \cite{rendle2009bpr}, SLIM \cite{ning2011slim}, and LorSLIM \cite{cheng2014lorslim}.
\begin{figure*}[!ht]
\centering
\subfloat[Delicious]{\includegraphics[width=.4\textwidth,height=4.5cm]{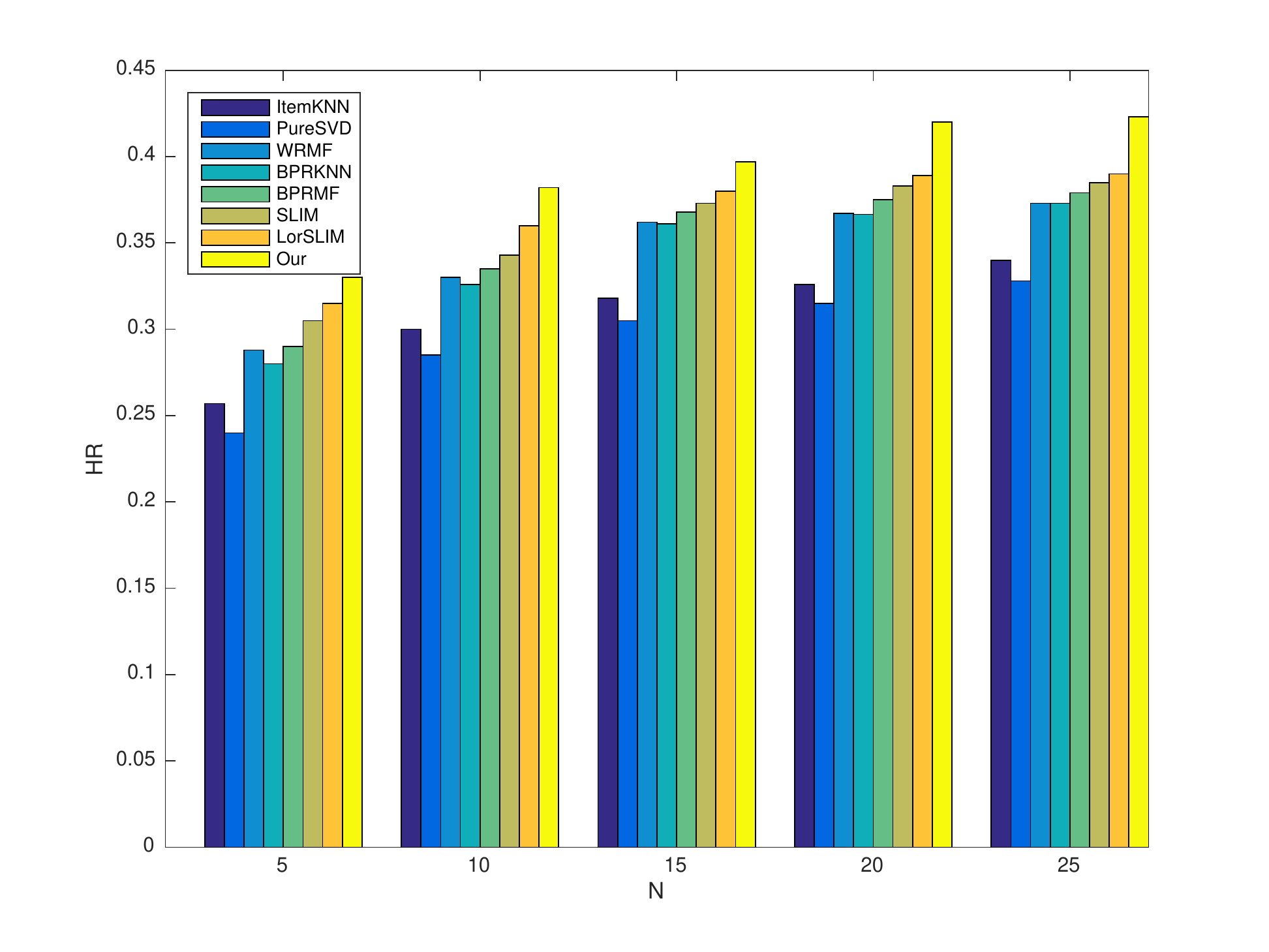}}
\subfloat[lastfm]{\includegraphics[width=.4\textwidth,height=4.5cm]{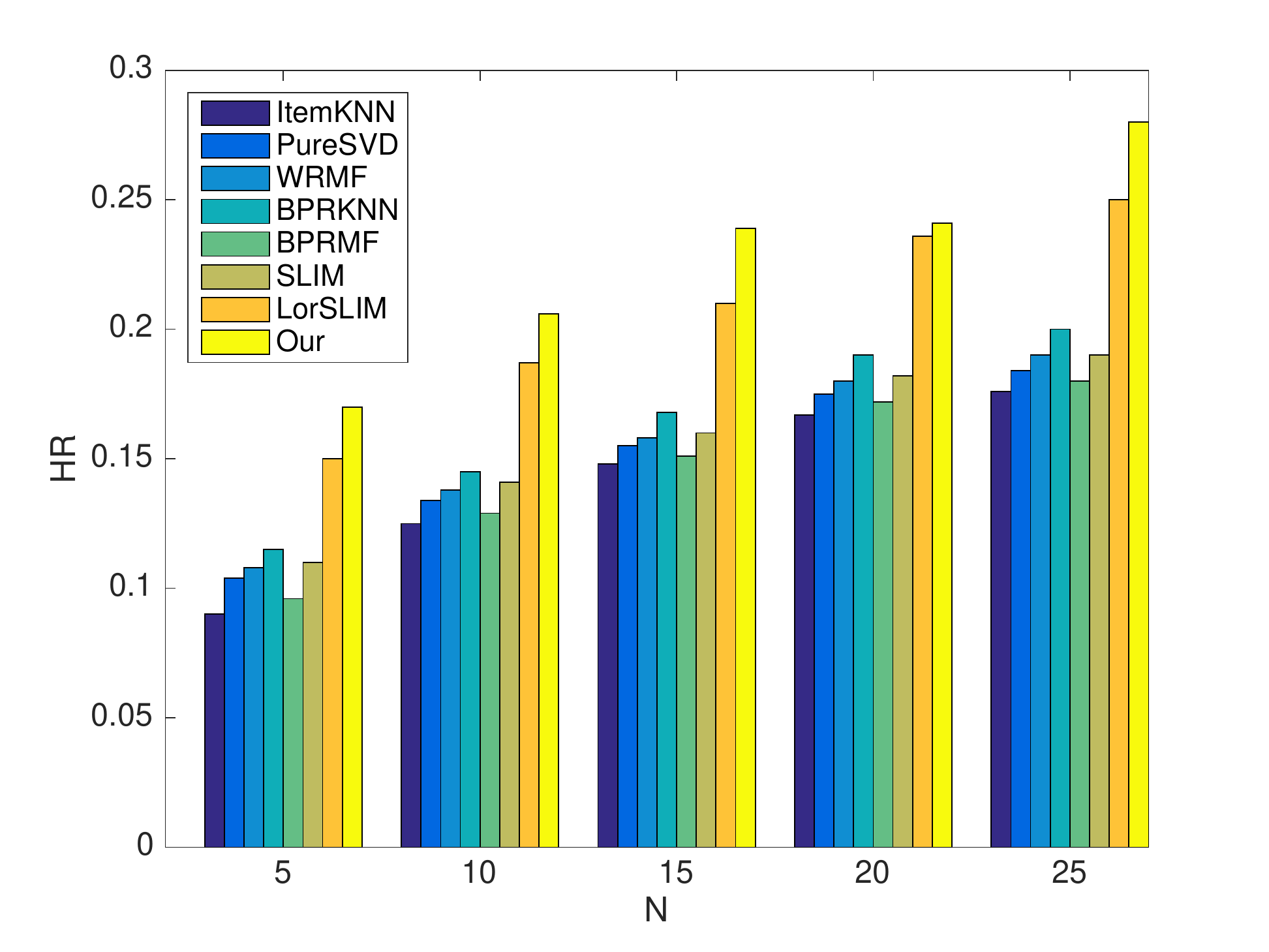}}\\
\subfloat[BX]{\includegraphics[width=.4\textwidth,height=4.5cm]{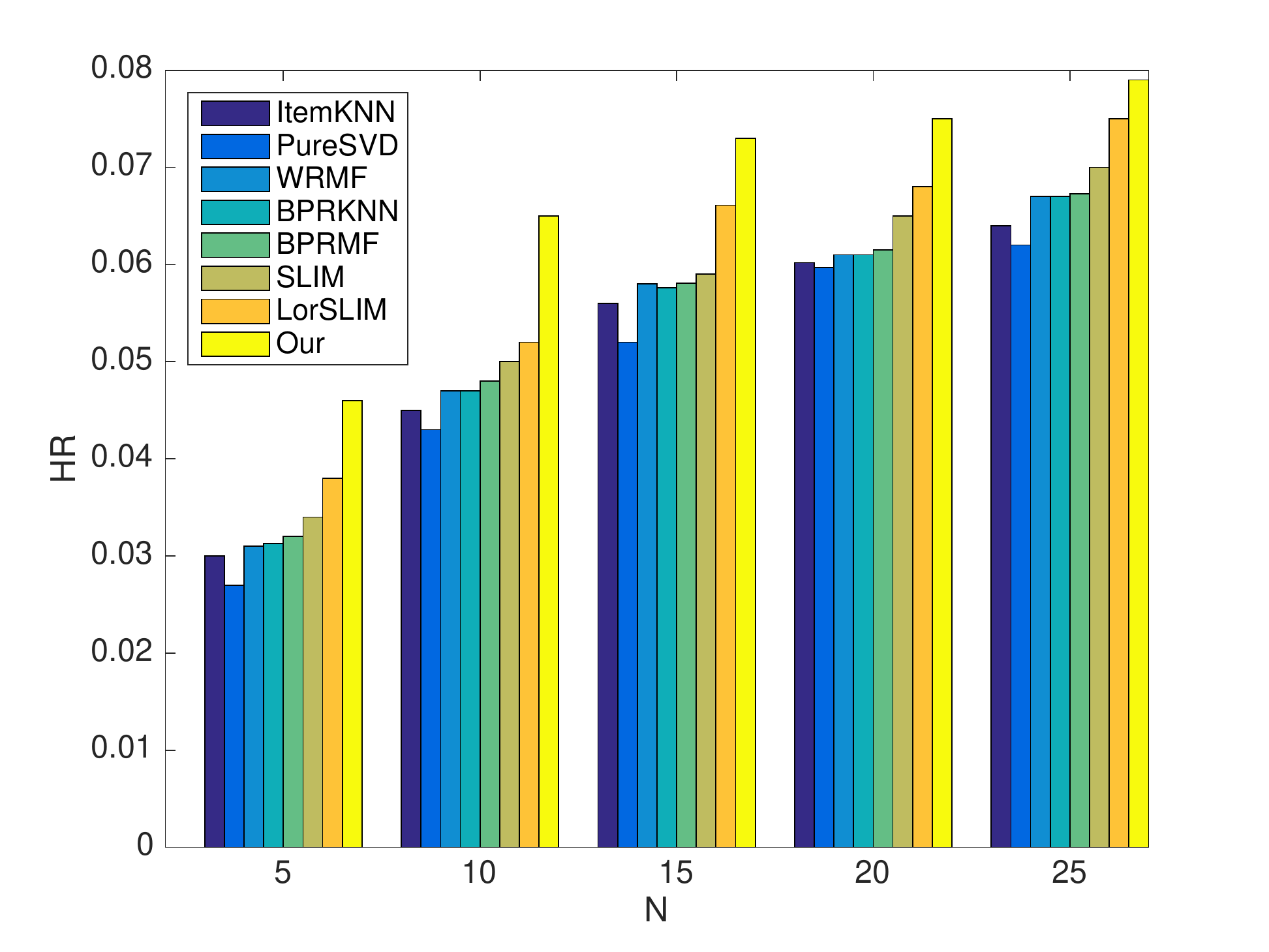}}
\subfloat[ML100K]{\includegraphics[width=.4\textwidth,height=4.5cm]{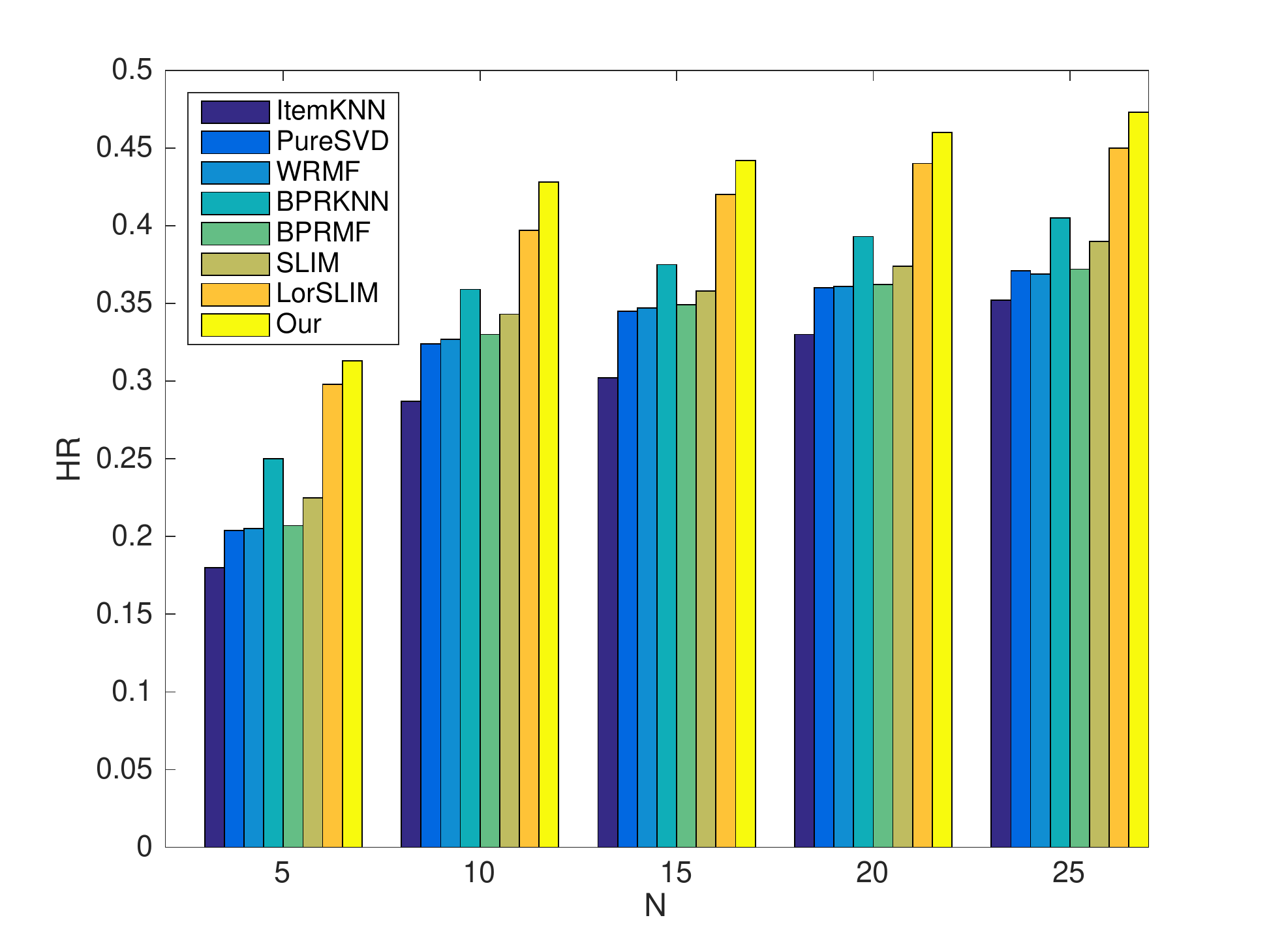}}\\
\subfloat[Netflix]{\includegraphics[width=.4\textwidth,height=4.5cm]{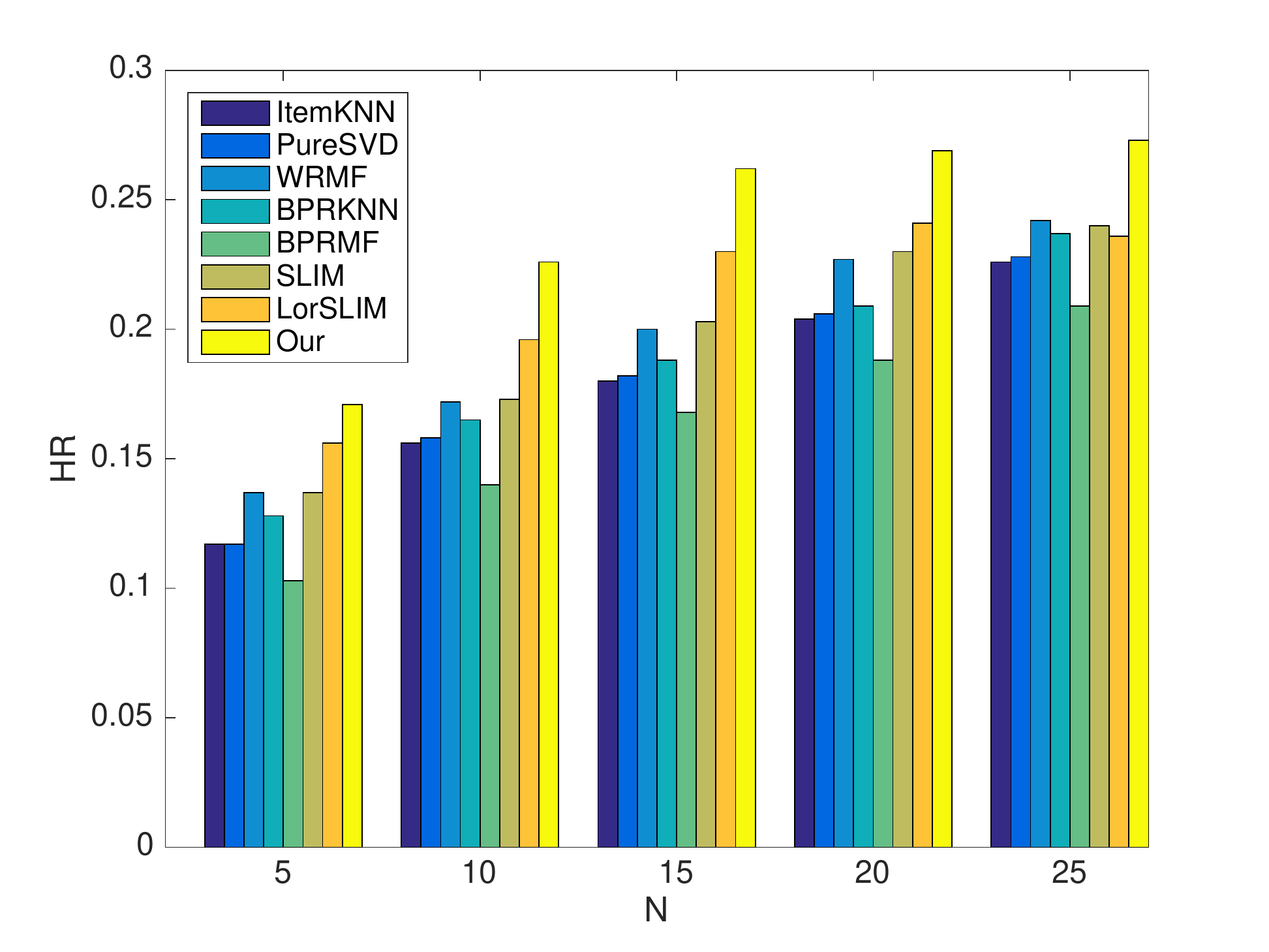}}
\subfloat[Yahoo]{\includegraphics[width=.4\textwidth,height=4.5cm]{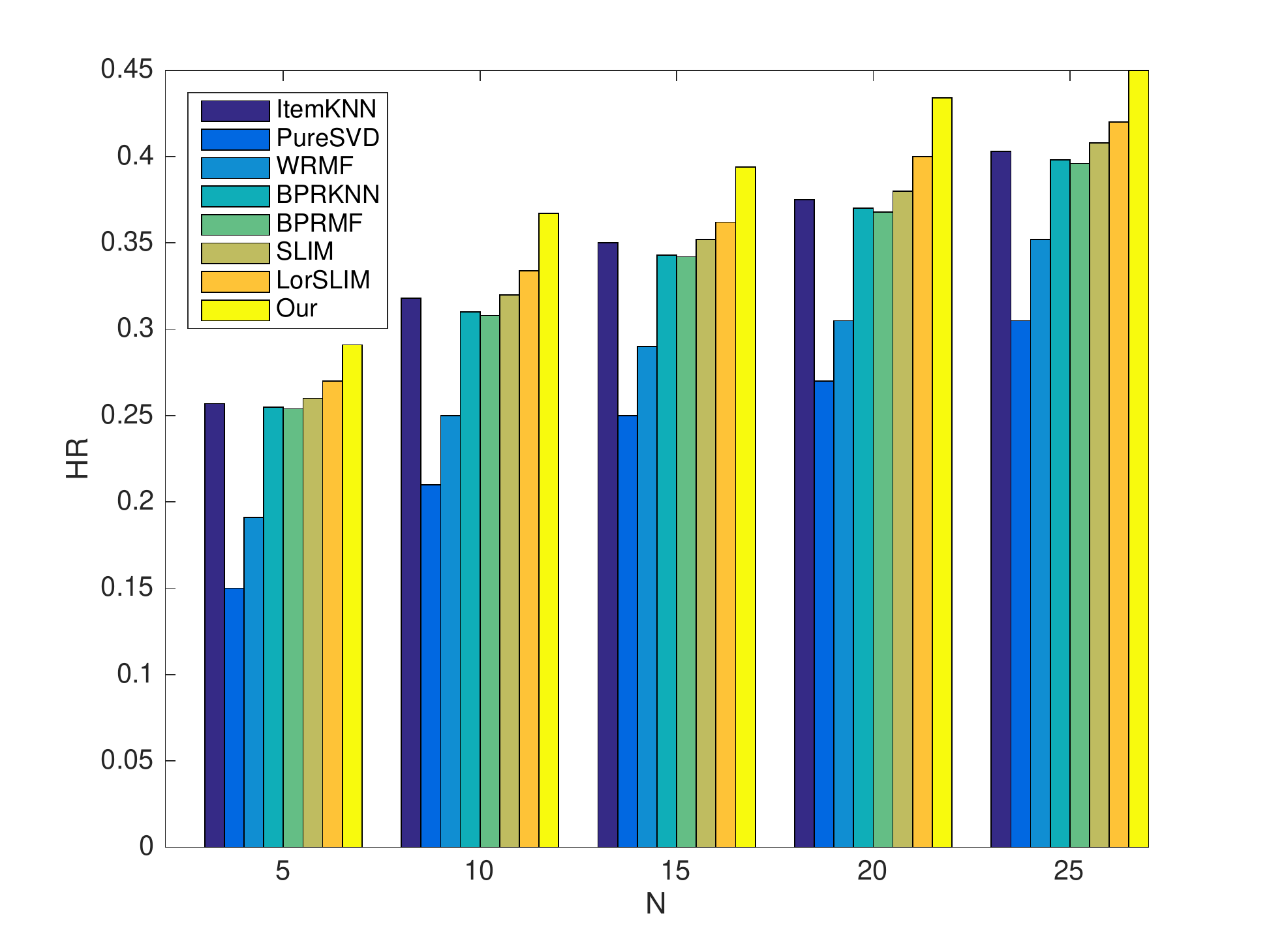}}
\caption{ Performance for Different Values of $N$.}
\label{fig:differentN}
\end{figure*}
\section{Experimental Results}
\subsection{Top-N recommendation performance}
We report the comparison results with other competing methods in Table \ref{tab:comp}.
The results show that our algorithm performs better than all the other methods across all the datasets \footnote{A bug is found, so the result in published version is updated. We apologize for any inconvenience caused.}. Specifically, in terms of HR, our method outperforms ItemKNN, PureSVD, WRMF, BPRKNN, BPRMF, SLIM and LorSLIM by 41\%, 48.14\%, 35.40\%, 28.69\%, 36.57\%, 26.26\%,  12.38\% on average, respectively, over all the six datasets; with respect to ARHR, ItemKNN, PureSVD, WRMF, BPRKNN, BPRMF, SLIM and LorSLIM are improved by 48.55\%, 60.38\%, 48.58\%, 37.14\%, 49.47\%, 31.94\%, 14.15\% on average, respectively. 

Among the seven state-of-the-art algorithms, LorSLIM is substantially better than the others. Moreover, SLIM is a little better than others except on lastfm and ML100K among the rest six methods. Then BPRKNN performs best among the remaining five methods on average.  Among the three MF-based models, BPRMF and WRMF have similar performance on most datasets and are much better than PureSVD on all datasets except on lastfm and ML100K.
\subsection{Recommendation for Different Top-N}

Figure \ref{fig:differentN} shows the performance achieved by the various methods for different values of $N$ for all six datasets. It demonstrates that the proposed method outperforms other algorithms in all scenarios. What is more, it is evident that the difference in performance between our approach and the other methods are consistently significant. It is interesting to note that LorSLIM, the second most competitive method, may be worse than some of the rest methods when $N$ is large.  
\subsection{Matrix Reconstruction}
We compare our method with LorSLIM by looking at how they reconstruct the user-item matrix. We take ML100K as an example, whose density is 6.30\% and the mean for those non-zero elements is 3.53. The reconstructed matrix from LorSLIM $\hat{X}=XW$ has a density of 13.61\%, whose non-zero values have a mean of 0.046. For those 6.30\% non-zero entries in $X$, $\hat{X}$ recovers 70.68\% of them and their mean value is 0.0665. This demonstrates that lots of information is lost. On the contrary, our approach fully preserves the original information thanks to the constraint condition in our model. Our method recovers all zero values with a mean of 0.554, which is much higher than 0.046. This suggests that our method recovers $X$ better than LorSLIM. This may explain the superior performance of our method.

\subsection{Parameter Tunning}
\begin{figure}[htbp]
\begin{center}
\includegraphics[width=.4\textwidth]{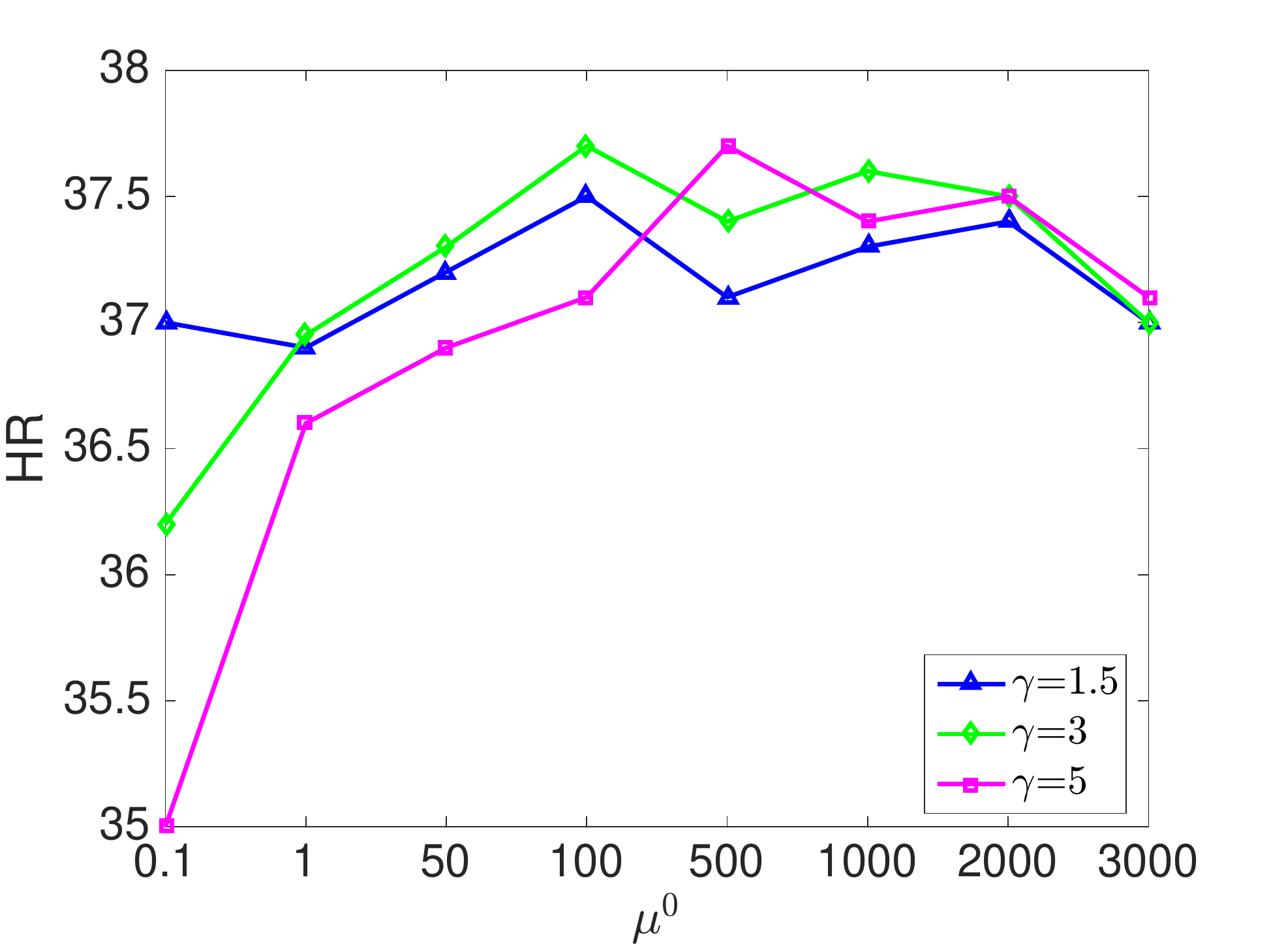}
\caption{ Influence of $\mu^0$ and $\gamma$ on HR for Delicious dataset.}
\label{fig:para}
\end{center}
\end{figure}
Although our model is parameter-free, we introduce the auxiliary parameter $\mu$ during the optimization. In alternating direction method of multipliers (ADMM) \cite{yang2013linearized}, $\mu$ is fixed and it is not easy to choose an optimal value to balance the computational cost. Thus, a dynamical $\mu$, increasing at a rate of $\gamma$, is preferred in real applications. $\gamma>1$ controls the convergence speed. The larger $\gamma$ is, the fewer iterations are to obtain the convergence, but meanwhile we may lose some precision. We show the effects of different initializations $\mu^0$ and $\gamma$ on HR on dataset Delicious in Figure \ref{fig:para}. It illustrates that our experimental results are not sensitive to them, which is reasonable since they are auxiliary parameters controlling mainly the convergence speed. In contrast, LorSLIM needs to tune four parameters, which are time consuming and not easy to operate. 
\subsection{Efficiency Analysis}

\begin{figure}[htbp]
\begin{center}
\includegraphics[width=.4\textwidth]{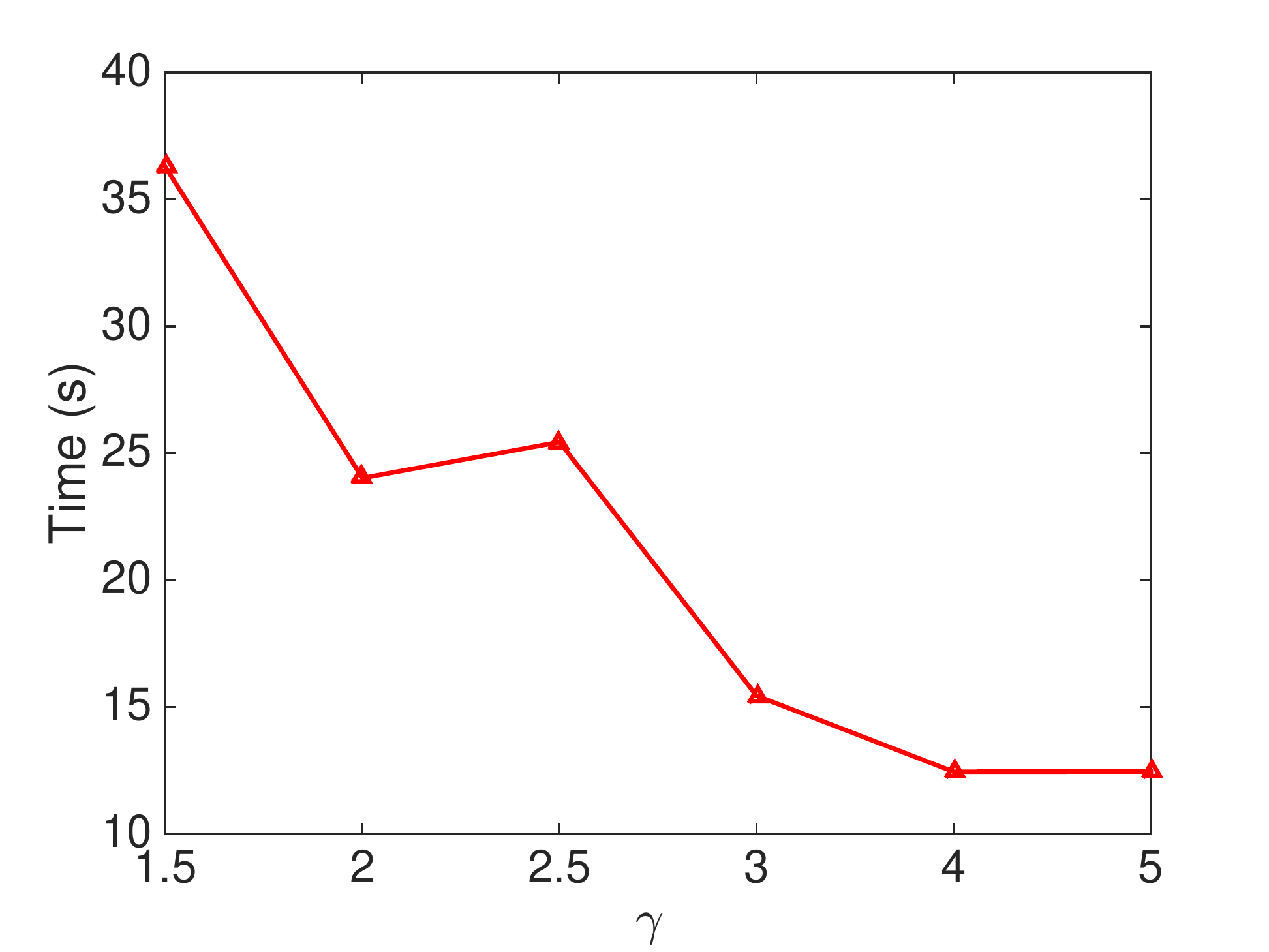}
\caption{ Influence of $\gamma$ on time.}
\label{fig:time}
\end{center}
\end{figure}
The time complexity of our algorithm is mainly from SVD. Exact SVD of a $m\times n$ matrix has a time complexity of $O(\min\{mn^2,m^2n\}$), in this paper we seek a low-rank matrix and thus only need a few principal singular vectors/values. Packages like PROPACK \cite{larsen2004propack} can compute a rank $k$ SVD with a cost of $O(min\{m^2k,n^2k\})$, which can be advantageous when $k\ll m,n$. In fact, our algorithm is much faster than LorSLIM. Among the six datasets, ML100K and lastfm datasets have the smallest and largest sizes, respectively. Our method needs 9s and 5080s, respectively, on these two datasets, while LorSLIM takes 617s and 32974s. The time is measured on the same machine with an Intel Xeon E3-1240 3.40GHz CPU that has 4 cores and 8GB memory, running Ubuntu and Matlab (R2014a). Furthermore, without losing too much accuracy, $\gamma$ can speed up our algorithm considerably. This is verified by Figure \ref{fig:time}, which shows the computational time of our method on Delicious with varying $\gamma$. 

\section{Conclusion}
In this paper, we present a matrix completion based method for the Top-N recommendation problem. The proposed method recovers the user-item matrix by solving a rank minimization problem. To better approximate the rank, a nonconvex function is applied. We conduct a comprehensive set of experiments on multiple datasets and compare its performance against that of other state-of-the-art Top-N recommendation algorithms. It turns out that our algorithm generates high quality recommendations, which improves the performance of the rest of methods considerably. This makes our approach usable in real-world scenarios.
\section{Acknowledgements}
This work is supported by US National Science Foundation Grants IIS 1218712. Q. Cheng is the corresponding author.  
\vfill\eject
\begin{quote}
\begin{small}
\bibliographystyle{aaai}
\bibliography{recom}

\begin{thebibliography}{}

\bibitem[\protect\citeauthoryear{Breese, Heckerman, and
  Kadie}{1998}]{breese1998empirical}
Breese, J.~S.; Heckerman, D.; and Kadie, C.
\newblock 1998.
\newblock Empirical analysis of predictive algorithms for collaborative
  filtering.
\newblock In {\em Proceedings of the Fourteenth conference on Uncertainty in
  artificial intelligence},  43--52.
\newblock Morgan Kaufmann Publishers Inc.

\bibitem[\protect\citeauthoryear{Cand{\`e}s and Recht}{2009}]{candes2009exact}
Cand{\`e}s, E.~J., and Recht, B.
\newblock 2009.
\newblock Exact matrix completion via convex optimization.
\newblock {\em Foundations of Computational mathematics} 9(6):717--772.

\bibitem[\protect\citeauthoryear{Cheng, Yin, and Yu}{2014}]{cheng2014lorslim}
Cheng, Y.; Yin, L.; and Yu, Y.
\newblock 2014.
\newblock Lorslim: Low rank sparse linear methods for top-n recommendations.
\newblock In {\em Data Mining (ICDM), 2014 IEEE International Conference on},
  90--99.
\newblock IEEE.

\bibitem[\protect\citeauthoryear{Cremonesi, Koren, and
  Turrin}{2010}]{cremonesi2010performance}
Cremonesi, P.; Koren, Y.; and Turrin, R.
\newblock 2010.
\newblock Performance of recommender algorithms on top-n recommendation tasks.
\newblock In {\em Proceedings of the fourth ACM conference on Recommender
  systems},  39--46.
\newblock ACM.

\bibitem[\protect\citeauthoryear{Degemmis, Lops, and
  Semeraro}{2007}]{degemmis2007content}
Degemmis, M.; Lops, P.; and Semeraro, G.
\newblock 2007.
\newblock A content-collaborative recommender that exploits wordnet-based user
  profiles for neighborhood formation.
\newblock {\em User Modeling and User-Adapted Interaction} 17(3):217--255.

\bibitem[\protect\citeauthoryear{Deshpande and
  Karypis}{2004}]{deshpande2004item}
Deshpande, M., and Karypis, G.
\newblock 2004.
\newblock Item-based top-n recommendation algorithms.
\newblock {\em ACM Transactions on Information Systems (TOIS)} 22(1):143--177.

\bibitem[\protect\citeauthoryear{Desrosiers and
  Karypis}{2011}]{desrosiers2011comprehensive}
Desrosiers, C., and Karypis, G.
\newblock 2011.
\newblock A comprehensive survey of neighborhood-based recommendation methods.
\newblock In {\em Recommender systems handbook}. Springer.
\newblock  107--144.

\bibitem[\protect\citeauthoryear{Fazel, Hindi, and Boyd}{2003}]{fazel2003log}
Fazel, M.; Hindi, H.; and Boyd, S.~P.
\newblock 2003.
\newblock Log-det heuristic for matrix rank minimization with applications to
  hankel and euclidean distance matrices.
\newblock In {\em American Control Conference, 2003. Proceedings of the 2003},
  volume~3,  2156--2162.
\newblock IEEE.

\bibitem[\protect\citeauthoryear{Good \bgroup et al\mbox.\egroup
  }{1999}]{good1999combining}
Good, N.; Schafer, J.~B.; Konstan, J.~A.; Borchers, A.; Sarwar, B.; Herlocker,
  J.; and Riedl, J.
\newblock 1999.
\newblock Combining collaborative filtering with personal agents for better
  recommendations.
\newblock In {\em AAAI/IAAI},  439--446.

\bibitem[\protect\citeauthoryear{Hu, Koren, and
  Volinsky}{2008}]{hu2008collaborative}
Hu, Y.; Koren, Y.; and Volinsky, C.
\newblock 2008.
\newblock Collaborative filtering for implicit feedback datasets.
\newblock In {\em Data Mining, 2008. ICDM'08. Eighth IEEE International
  Conference on},  263--272.
\newblock IEEE.

\bibitem[\protect\citeauthoryear{Kang and Cheng}{2015}]{kangICDM}
Kang, Zhao ang~Peng, C., and Cheng, Q.
\newblock 2015.
\newblock Robust pca via nonconvex rank approximation.
\newblock In {\em Data Mining (ICDM), 2015 IEEE International Conference on},
  211--220.
\newblock IEEE.

\bibitem[\protect\citeauthoryear{Kang \bgroup et al\mbox.\egroup
  }{2015}]{kang2015logdet}
Kang, Z.; Peng, C.; Cheng, J.; and Cheng, Q.
\newblock 2015.
\newblock Logdet rank minimization with application to subspace clustering.
\newblock {\em Computational intelligence and neuroscience} 2015:68.

\bibitem[\protect\citeauthoryear{Kang, Peng, and Cheng}{2015a}]{kang2015robust}
Kang, Z.; Peng, C.; and Cheng, Q.
\newblock 2015a.
\newblock Robust subspace clustering via robust subspace clustering via
  smoothed rank approximation.
\newblock {\em SIGNAL PROCESSING LETTERS, IEEE} 22(11):2088--2092.

\bibitem[\protect\citeauthoryear{Kang, Peng, and
  Cheng}{2015b}]{kangcikm2015robust}
Kang, Z.; Peng, C.; and Cheng, Q.
\newblock 2015b.
\newblock Robust subspace clustering via tighter rank approximation.
\newblock {\em ACM CIKM'15}.

\bibitem[\protect\citeauthoryear{Larsen}{2004}]{larsen2004propack}
Larsen, R.~M.
\newblock 2004.
\newblock Propack-software for large and sparse svd calculations.
\newblock {\em Available online. URL http://sun. stanford. edu/rmunk/PROPACK}
  2008--2009.

\bibitem[\protect\citeauthoryear{Lee \bgroup et al\mbox.\egroup
  }{2014}]{lee2014local}
Lee, J.; Bengio, S.; Kim, S.; Lebanon, G.; and Singer, Y.
\newblock 2014.
\newblock Local collaborative ranking.
\newblock In {\em Proceedings of the 23rd international conference on World
  wide web},  85--96.
\newblock ACM.

\bibitem[\protect\citeauthoryear{Li and Za{\"\i}ane}{2004}]{li2004combining}
Li, J., and Za{\"\i}ane, O.~R.
\newblock 2004.
\newblock Combining usage, content, and structure data to improve web site
  recommendation.
\newblock In {\em E-Commerce and Web Technologies}. Springer.
\newblock  305--315.

\bibitem[\protect\citeauthoryear{Linden, Smith, and
  York}{2003}]{linden2003amazon}
Linden, G.; Smith, B.; and York, J.
\newblock 2003.
\newblock Amazon. com recommendations: Item-to-item collaborative filtering.
\newblock {\em Internet Computing, IEEE} 7(1):76--80.

\bibitem[\protect\citeauthoryear{Lu \bgroup et al\mbox.\egroup
  }{2014}]{lu2014generalized}
Lu, C.; Tang, J.; Yan, S.; and Lin, Z.
\newblock 2014.
\newblock Generalized nonconvex nonsmooth low-rank minimization.
\newblock In {\em Computer Vision and Pattern Recognition (CVPR), 2014 IEEE
  Conference on},  4130--4137.
\newblock IEEE.

\bibitem[\protect\citeauthoryear{Melville, Mooney, and
  Nagarajan}{2002}]{melville2002content}
Melville, P.; Mooney, R.~J.; and Nagarajan, R.
\newblock 2002.
\newblock Content-boosted collaborative filtering for improved recommendations.
\newblock In {\em AAAI/IAAI},  187--192.

\bibitem[\protect\citeauthoryear{Ning and Karypis}{2011}]{ning2011slim}
Ning, X., and Karypis, G.
\newblock 2011.
\newblock Slim: Sparse linear methods for top-n recommender systems.
\newblock In {\em Data Mining (ICDM), 2011 IEEE 11th International Conference
  on},  497--506.
\newblock IEEE.

\bibitem[\protect\citeauthoryear{Pan \bgroup et al\mbox.\egroup
  }{2008}]{pan2008one}
Pan, R.; Zhou, Y.; Cao, B.; Liu, N.~N.; Lukose, R.; Scholz, M.; and Yang, Q.
\newblock 2008.
\newblock One-class collaborative filtering.
\newblock In {\em Data Mining, 2008. ICDM'08. Eighth IEEE International
  Conference on},  502--511.
\newblock IEEE.

\bibitem[\protect\citeauthoryear{Recht, Xu, and
  Hassibi}{2008}]{recht2008necessary}
Recht, B.; Xu, W.; and Hassibi, B.
\newblock 2008.
\newblock Necessary and sufficient conditions for success of the nuclear norm
  heuristic for rank minimization.
\newblock In {\em Decision and Control, 2008. CDC 2008. 47th IEEE Conference
  on},  3065--3070.
\newblock IEEE.

\bibitem[\protect\citeauthoryear{Rendle \bgroup et al\mbox.\egroup
  }{2009}]{rendle2009bpr}
Rendle, S.; Freudenthaler, C.; Gantner, Z.; and Schmidt-Thieme, L.
\newblock 2009.
\newblock Bpr: Bayesian personalized ranking from implicit feedback.
\newblock In {\em Proceedings of the Twenty-Fifth Conference on Uncertainty in
  Artificial Intelligence},  452--461.
\newblock AUAI Press.

\bibitem[\protect\citeauthoryear{Ricci, Rokach, and
  Shapira}{2011}]{ricci2011introduction}
Ricci, F.; Rokach, L.; and Shapira, B.
\newblock 2011.
\newblock {\em Introduction to recommender systems handbook}.
\newblock Springer.

\bibitem[\protect\citeauthoryear{Shi and Yu}{2011}]{shi2011limitations}
Shi, X., and Yu, P.~S.
\newblock 2011.
\newblock Limitations of matrix completion via trace norm minimization.
\newblock {\em ACM SIGKDD Explorations Newsletter} 12(2):16--20.

\bibitem[\protect\citeauthoryear{Srebro and
  Salakhutdinov}{2010}]{srebro2010collaborative}
Srebro, N., and Salakhutdinov, R.~R.
\newblock 2010.
\newblock Collaborative filtering in a non-uniform world: Learning with the
  weighted trace norm.
\newblock In {\em Advances in Neural Information Processing Systems},
  2056--2064.

\bibitem[\protect\citeauthoryear{Yang and Yuan}{2013}]{yang2013linearized}
Yang, J., and Yuan, X.
\newblock 2013.
\newblock Linearized augmented lagrangian and alternating direction methods for
  nuclear norm minimization.
\newblock {\em Mathematics of Computation} 82(281):301--329.

\bibitem[\protect\citeauthoryear{Yu \bgroup et al\mbox.\egroup
  }{2009}]{yu2009fast}
Yu, K.; Zhu, S.; Lafferty, J.; and Gong, Y.
\newblock 2009.
\newblock Fast nonparametric matrix factorization for large-scale collaborative
  filtering.
\newblock In {\em Proceedings of the 32nd international ACM SIGIR conference on
  Research and development in information retrieval},  211--218.
\newblock ACM.

\bibitem[\protect\citeauthoryear{Zhong \bgroup et al\mbox.\egroup
  }{2015}]{zhong2015nonconvex}
Zhong, X.; Xu, L.; Li, Y.; Liu, Z.; and Chen, E.
\newblock 2015.
\newblock A nonconvex relaxation approach for rank minimization problems.
\newblock In {\em Twenty-Ninth AAAI Conference on Artificial Intelligence}.

\end{thebibliography}
\end{small}
\end{quote}

\end{document}